\documentclass[aps,prl,twocolumn]{revtex4}
\usepackage{amsmath,amssymb,xcolor}
\input epsf
\usepackage{verbatim}
\usepackage{graphicx}
\usepackage{bm}
\def\pmb#1{\setbox0=\hbox{#1}
\kern-.025em\copy0\kern-\wd0 \kern-.05em\copy0\kern-\wd0
\kern-.025em\raise.0433em\box0}

\newcommand{\beq}{\begin{equation}}
\newcommand{\eeq}{\end{equation}}
\newcommand{\ba}{\begin{eqnarray}}
\newcommand{\ea}{\end{eqnarray}}

\usepackage[caption=false]{subfig}

\definecolor{colorcb}{rgb}{0 0.4470 0.7410}

\definecolor{colorcbq}{rgb}{0.4660 0.6740 0.1880}

\definecolor{rouge}{rgb}{1,0,0}
\definecolor{bleu}{rgb}{0,0,1}
\definecolor{vert}{rgb}{0,0.5,0}
\newcommand{\blue}[1]{{\color{bleu}#1}}

\input epsf

\graphicspath{{Figures/}}

\usepackage[english]{babel}
\begin{document}

\title[]{Time-domain investigation of a cylindrical acoustic external cloak}
\author{ S. Guenneau$^{1}$, B. Lombard$^{2}$, C. Bellis$^{2}$}
\affiliation{$^1$ UMI 2004 Abraham de Moivre-CNRS, Imperial College London, SW7 2AZ, United Kingdom}
\affiliation{$^2$ Aix Marseille Univ, CNRS, Centrale Marseille, LMA, Marseille, France}

\begin{abstract}
Space folding techniques based on non-monotonic transforms lead to a new class of cylindrical isotropic acoustic cloaks with a constant negative density and a spatially varying negative bulk modulus. We consider an external cloak consisting of a core with a positive definite density matrix and a positive compressibility, and a shell with simultaneously negative density and compressibility. Such a core-shell resonant system creates a virtual folded region outside the shell. To handle such negative physical parameters in the time-domain, a two-step strategy is used: (i) assuming resonant (Drude-type) effective parameters in the frequency-domain; (ii) returning to the time-domain by applying the formalism of the auxiliary fields. We numerically show that, at the designed central frequency, scattering of a cylindrical pressure wave incident upon a finite set of small rigid obstacles is drastically reduced after a lapse of time, when they are placed in the close neighborhood of the external cloak. However, at short times, the external cloak behaves rather like a superscatterer: the cloak itself scatters more than the set of scatterers.

\pacs{41.20.Jb,42.25.Bs,42.70.Qs,43.20.Bi,43.25.Gf}

\end{abstract}
\maketitle


It is by now well established that one can cloak objects surrounded by specially designed anisotropic heterogeneous \cite{cloakall,ulf06,oref0,greenleaf03}, or plasmonic \cite{alu05}, shells. It is even possible to cloak objects using some active sources \cite{miller06,vasquez09}. In this Letter, our objective is to analyze the possibility to reduce the scattering of small objects placed outside a core-shell system with sign-shifting heterogeneous (yet isotropic) coefficients.

Here, we focus on the simple case of scalar pressure waves propagating in a non-viscous fluid which are governed by the acoustic equations:
\begin{equation}
\left\{
\begin{array}{l}
\rho(\bm{x})\,\partial_t \bm{v}(\bm{x},t)+\bm{\nabla} p(\bm{x},t)={\bf0},\\ [4pt]
\kappa^{-1}(\bm{x})\,\partial_t p(\bm{x},t)+\bm{\nabla}\cdot\bm{v}(\bm{x},t)=0,
\end{array}
\right.
\label{wave1} 
\end{equation}
where $\bm{v}$ is the acoustic velocity, $p$ the acoustic pressure, $\rho$ the mass density (kg.m$^{-3}$) and $\kappa$ the bulk modulus (Pa) of the fluid, all of which depend upon the space and time variables $\bm{x}=(x_1,x_2)\in\mathbb{R}^2$ and $t\geq 0$, with $\bm{\nabla}=(\partial/\partial x_1,\partial/\partial x_2)^T$. 

Inspired by earlier work on anomalous resonance and cloaking with resonant core-shell systems based on the electrostatic \cite{oref1,oref7,oref10,oref9,oref15,bouchitte10,oref17} and electromagnetic \cite{oref11,oref13,oref16} counterparts of equation \eqref{wave1} in the time-harmonic regime, we consider the following set of radially symmetric parameters for an external acoustic cloak with core and shell radii $R_c<R_s$ (see Fig. \ref{FigCloak}):
\begin{equation}
\begin{aligned}
& \rho(\bm{x};\omega_0)=\rho_e
\left\{
\begin{aligned}
& +1 && 0\leq |\bm{x}|<R_c,\\
& -1 && R_c<|\bm{x}| \leq R_s,\\
& +1 && R_s<|\bm{x}|,
\end{aligned}
\right.\\[1mm]
& \kappa^{-1}(\bm{x};\omega_0)=\kappa_e^{-1}
\left\{
\begin{aligned}
& +\left({R_s}/{R_c}\right)^4 && 0\leq |\bm{x}|<R_c,\\
& -\left({R_s}/{|\bm{x}|}\right)^4 && R_c<|\bm{x}| \leq R_s,\\
& +1 && R_s<|\bm{x}|,
\end{aligned}
\right.
\end{aligned}
\label{ParamR}
\end{equation}
with $|\bm{x}|=\sqrt{x_1^2+x_2^2}$ and $\rho_e,\kappa_e>0$, while the central circular frequency parameter $\omega_0$ emphasizes that \eqref{ParamR} pertains to the wave equation in the frequency-domain.
{ We note that these material parameters can be inferred from earlier work on the analogous transverse electromagnetic problem \cite{oref13,oref16} where the electric permittivity $\varepsilon$ plays the role of $\rho$ and the magnetic permeability $\mu$ that of $\kappa^{-1}$ \cite{cummer08}, or vice versa, depending upon the light polarization.
}
The specific form of \eqref{ParamR} achieves impedance-matching at the core and shell interfaces and it has been related in \cite{oref9,oref13,oref16} (albeit in an electrostatic/time-harmonic electromagnetic configuration) to the coordinate transform $\phi:\bm{x}\mapsto\bm{x}'$ defined as:
\begin{equation}	       
  |\bm{x}'|=\phi(|\bm{x}|)=  \left\{ \begin{aligned}
& R_s^2 |\bm{x}|/R_c^2  && \text{for }  |\bm{x}|\le R_c, \\  
& R_s^2/|\bm{x}|	  && \text{for }  R_c\blue{<} |\bm{x}| \le R_s, \\
&|\bm{x}| 	&& \text{for }  |\bm{x}|>R_s.
\end{aligned}
\right.
\label{shad-eq1}
\end{equation}
Unlike in usual transformation acoustics where the mass density is mapped, in general, to a rank-2 tensor while the bulk modulus remains a scalar \cite{norris08}, we note that 
the use of the coordinate transform \eqref{shad-eq1} leads to a transformed density $\rho'(\bm{x}';\omega_0)$ and a bulk modulus $\kappa'(\bm{x}';\omega_0)$ which both remain scalar functions in each subdomain. Geometrically, $\phi$ maps the core disc $\mathcal{C}=D(0,R_c)$ onto $D(0,R_*)$, with $R_*=R_s^2/R_c$, and \emph{folds} the shell annulus $\mathcal{S}=D(0,R_s)-D(0,R_c)$ onto $D(0,R_s)-D(0,R_*)$, see Figure~\ref{FigCloak}. Note that $\phi$ in \eqref{shad-eq1} is a non-bijective (and non-smooth) function, which is a genuine feature of space folding.

\begin{figure}[h!]
\includegraphics[width=0.48\textwidth]{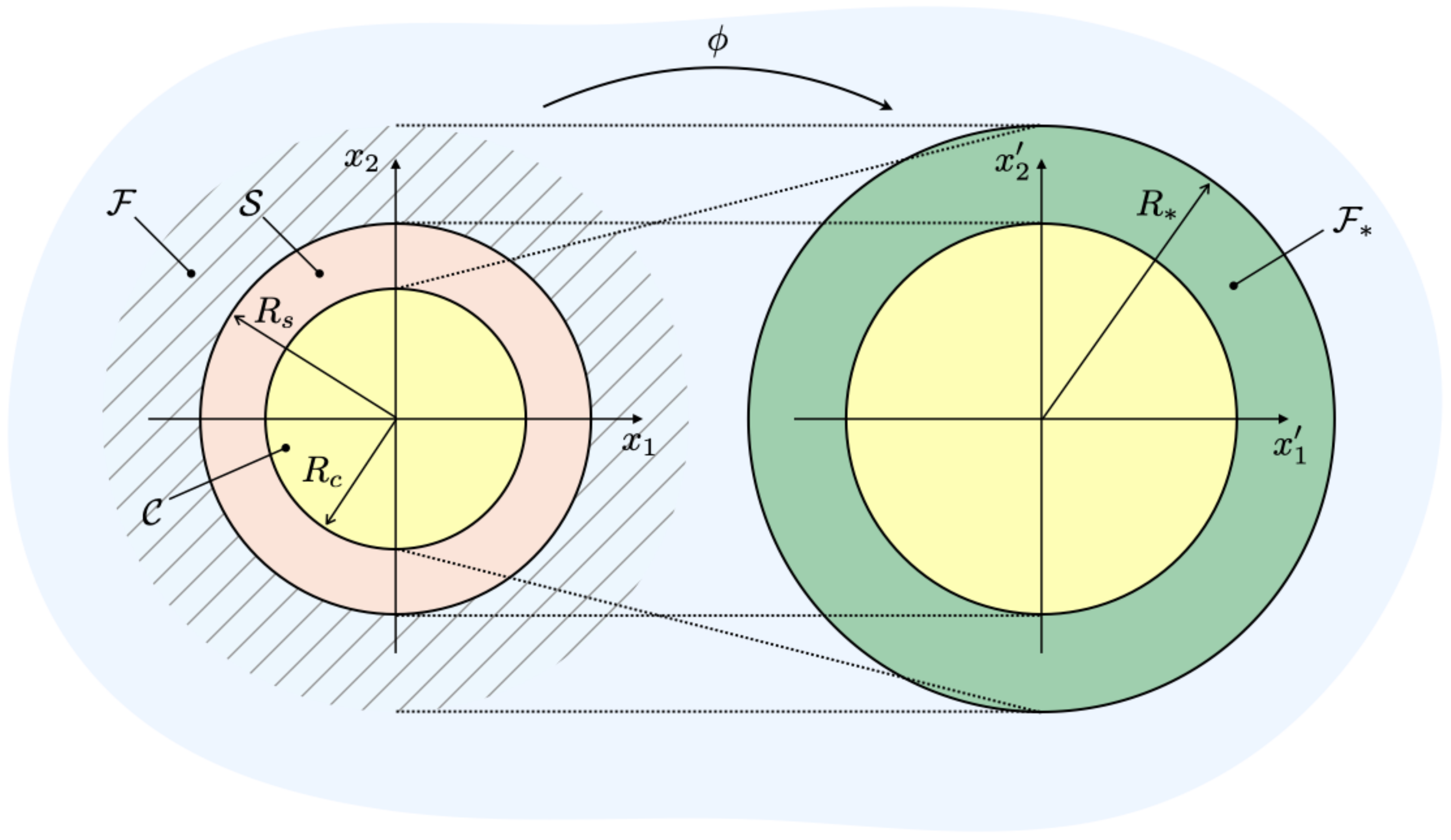}
\caption{Left: schematics of the core-shell system in the physical domain with the external cloaking region $\mathcal{F}$ (dashed area). Right: the transformed domain with the virtual folded region $\mathcal{F}_*$ (in green).}
\vspace{-1mm}
\label{FigCloak}
\end{figure}

{Thanks to $\phi$ in (\ref{shad-eq1}), we thus create a folding zone in space, namely the annulus $\mathcal{F}_*$ of width $| R_s-R_* |$,} which corresponds to the so-called external cloaking region $\mathcal{F}$ in the physical space \cite{oref13} (with $\mathcal{F}_*\equiv \mathcal{F}$). Physically, this means that the acoustic field outside the region of radius $R_*$\,should not sense what is inside this region, and this has been confirmed by a mathematical analysis based on the Helmholtz equation in the time-harmonic regime \cite{oref16}. Interestingly, the core-shell system \eqref{shad-eq1} can also be viewed as a folded version of the Veselago-Pendry lens \cite{neglens,pendry00,milton05}, via the concept of complementary media \cite{pendry02}, which is reminiscent of space-folding. Actually, it has been numerically shown in the time-harmonic regime that if one considers a finite size object outside the external cloak and {a ``space folded'' copy of the object consisting of opposite sign electromagnetic parameters} inside the shell, then this makes a perfectly transparent electromagnetic {core-shell} system at any frequency \cite{lai09}. 

\begin{figure}[htbp]
\centering
\subfloat[$t=35.6$\,ms]{\hspace{-5mm}\includegraphics[scale=0.25]{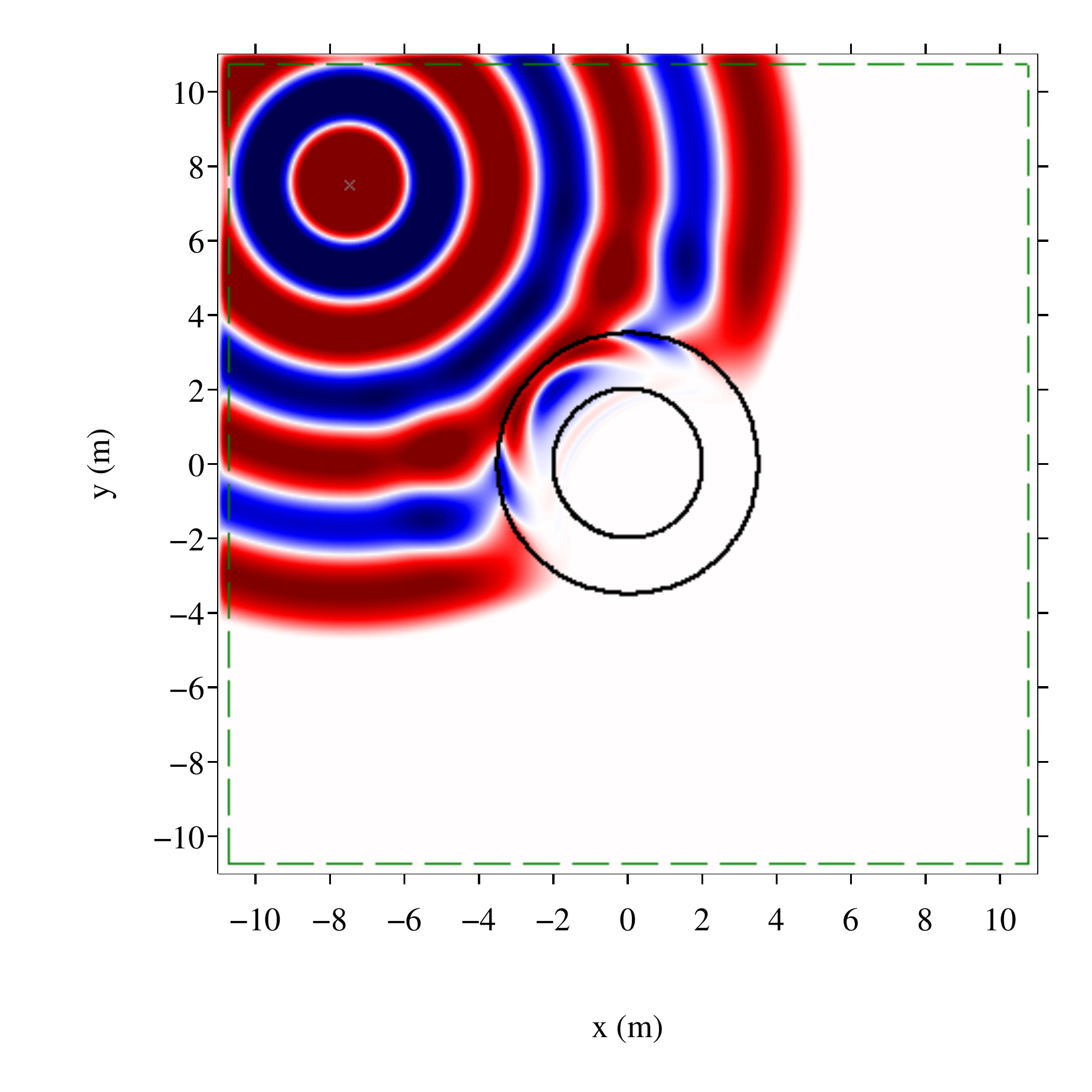}}
\subfloat[$t=71.2$\,ms]{\includegraphics[scale=0.25]{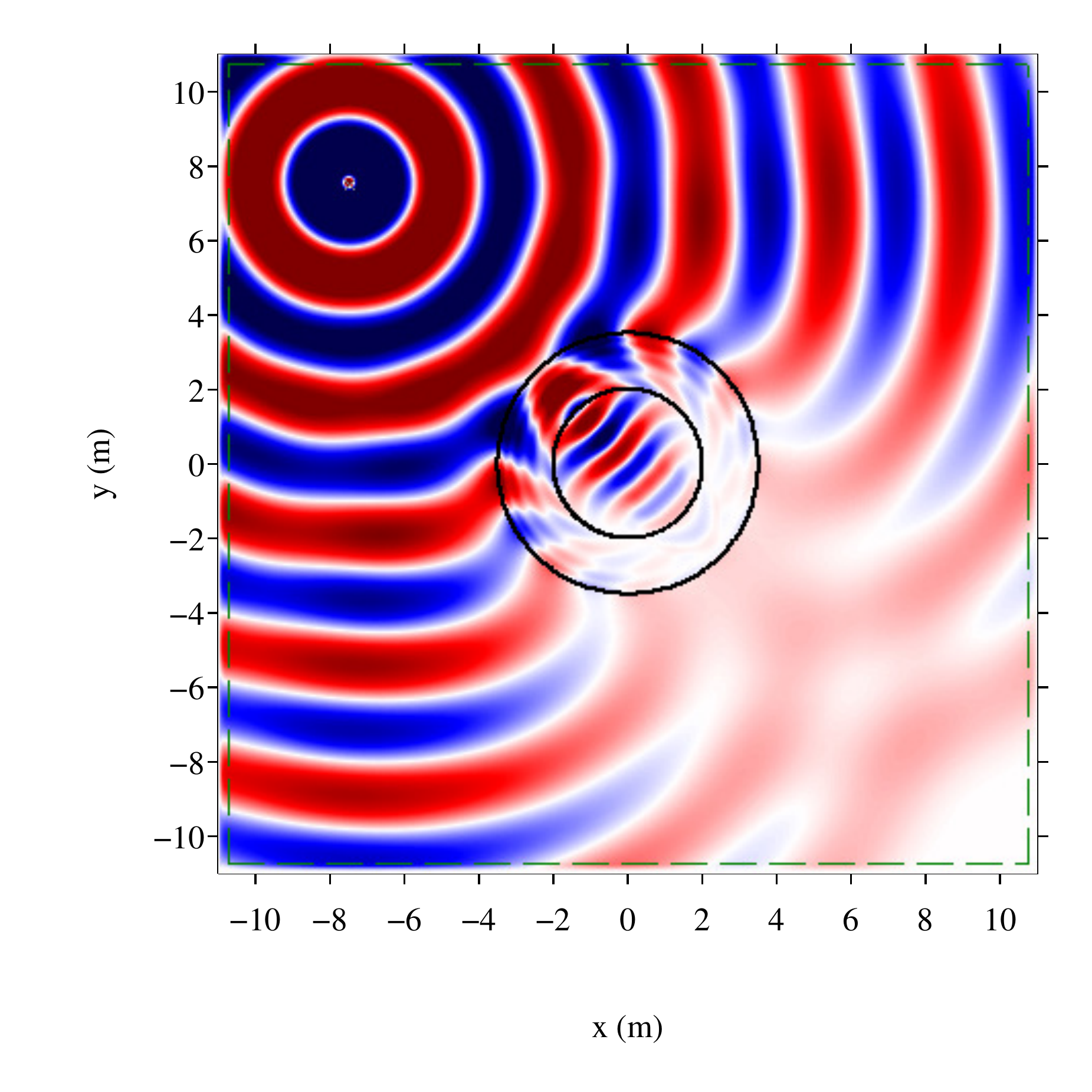}}\\[-4mm]
\subfloat[$t=1.424$\,s]{\hspace{-5mm}\includegraphics[scale=0.25]{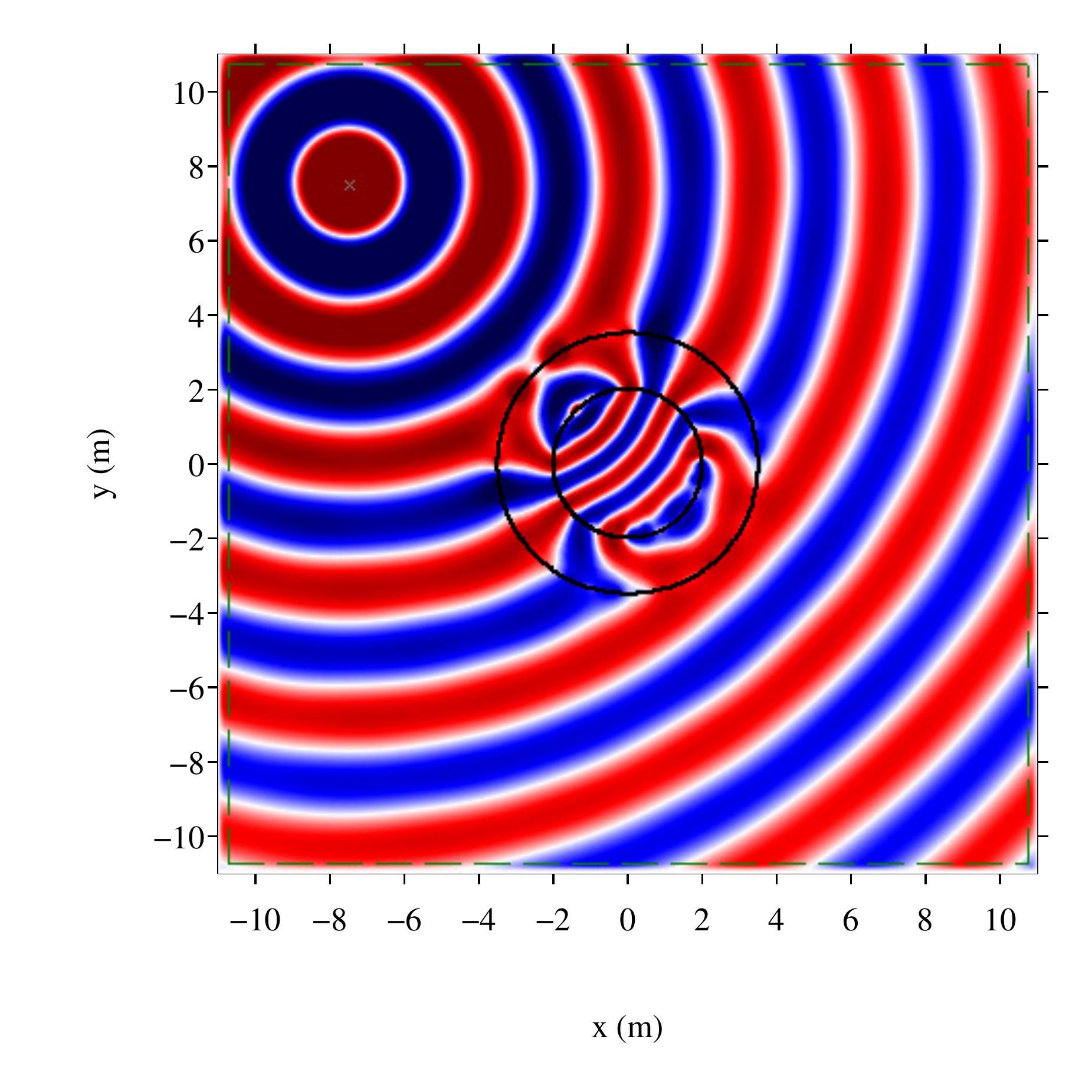}}
\subfloat[$t=1.424$\,s close-up]{\includegraphics[scale=0.25]{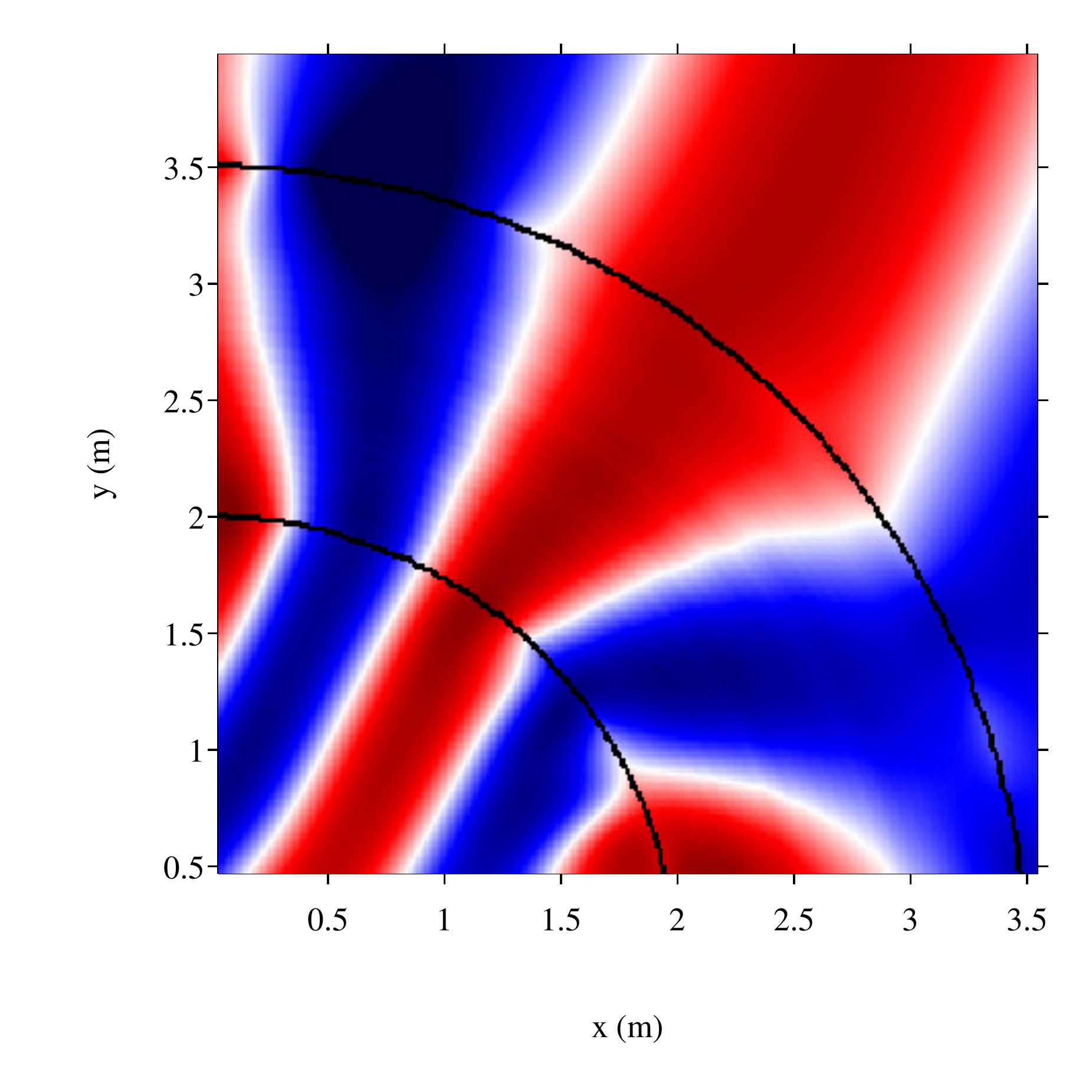}}
\vspace{-2mm}
\caption{\label{FigCloak}Snapshots of the pressure field $p$ at different times for the cloak only. Panel (d) is a close-up of (c) highlighting the negatively refracted waves in the {isotropic} shell.}
\end{figure} 

However, despite all these works, the transient behavior of waves propagating across the core-shell system \eqref{ParamR} remained an open question until now, and we would like to address it through time-domain numerical simulations. Indeed, we can then make use of the powerful formalism of auxiliary field for electromagnetic \cite{gralak10} and acoustic \cite{Bellis19} waves propagating within sign-shifting media.

In the core $\mathcal{C}$ and in the matrix $\mathbb{R}^2\setminus(\mathcal{C}\cup\mathcal{S})$, the physical parameters $\rho(\bm{x};\omega_0)$ and $\kappa(\bm{x};\omega_0)$ in \eqref{ParamR} are both positive, which meets the standard assumption for a mathematically, and physically, {well-posed problem}. On the other hand, in the shell $\mathcal{S}$, both $\rho(\bm{x};\omega_0)$ and $\kappa(\bm{x};\omega_0)$ are negative, which is non-standard and requires special treatment in order to investigate external cloaking in the time-domain. To do so, we consider that such negative parameters can be realized, at the given frequency $\omega_0$, as the macroscopic (homogenized) behavior in the time-harmonic regime of a resonant micro-structured medium (which we do not intend to describe here) filling this region of space. Resonant micro-structures generally lead to frequency-dependent effective parameters, a simple instance of which is the following Drude-like model relevant to the frequency-domain (Helmholtz) wave equation:
\begin{equation}
\rho(\bm{x},\omega)=\rho(\bm{x};\omega_0)\,g(\omega), \ 
\kappa^{-1}(\bm{x},\omega)=\kappa^{-1}(\bm{x};\omega_0)\,g(\omega),
\label{Drude1}
\end{equation}
for $\bm{x}\in\mathcal{S}$ and with
\begin{equation}
g(\omega)= 1-\left({\Omega_0}/{\omega}\right)^2.
\label{Drude2}
\end{equation} 
Choosing $\Omega_0=\sqrt{2}\,\omega_0$ yields $g(\omega_0)=-1$. To handle such frequency-dependent parameters in the time-domain wave equation, we make use of the formalism of auxiliary fields \cite{Bellis19}. By introducing locally the fields $\bm{w}$ and $q$, the equations of acoustics in the shell $\mathcal{S}$ are rewritten as the augmented system:
\begin{equation}
\left\{
\begin{array}{l}
\rho_e\,\partial_t \bm{v}+\rho_e\,\Omega_0^2\,\partial_t\bm{w}+\bm{\nabla} p={\bf0},\\ [4pt]
\kappa^{-1}(\bm{x};\omega_0)\,\partial_t p+\kappa^{-1}(\bm{x};\omega_0)\,\Omega_0^2\,\partial_t q+\bm{\nabla}\cdot\bm{v}=0,\\ [4pt]
\partial^2_{tt}\bm{w}=\bm{v},\\ [4pt]
\partial^2_{tt}q=p,
\end{array}
\right.
\label{Auxiliary} 
\end{equation}
where we emphasize only the variable dependency of the constitutive parameters. Equations \eqref{Auxiliary} are completed with \eqref{wave1} in the core $\mathcal{C}$ and in the matrix $\mathbb{R}^2\setminus(\mathcal{C}\cup\mathcal{S})$ with the corresponding parameters in \eqref{ParamR}. In addition, we assume the usual continuity conditions of $p$ and $\bm{v}\cdot \bm{n}$ at the core and shell interfaces, with $\bm{n}$ the unit normal vector at the interfaces, and suitable outgoing wave conditions at infinity.

\begin{figure}[htbp]
\centering
\subfloat[$t=35.6$\,ms]{\hspace{-5mm}\includegraphics[scale=0.25]{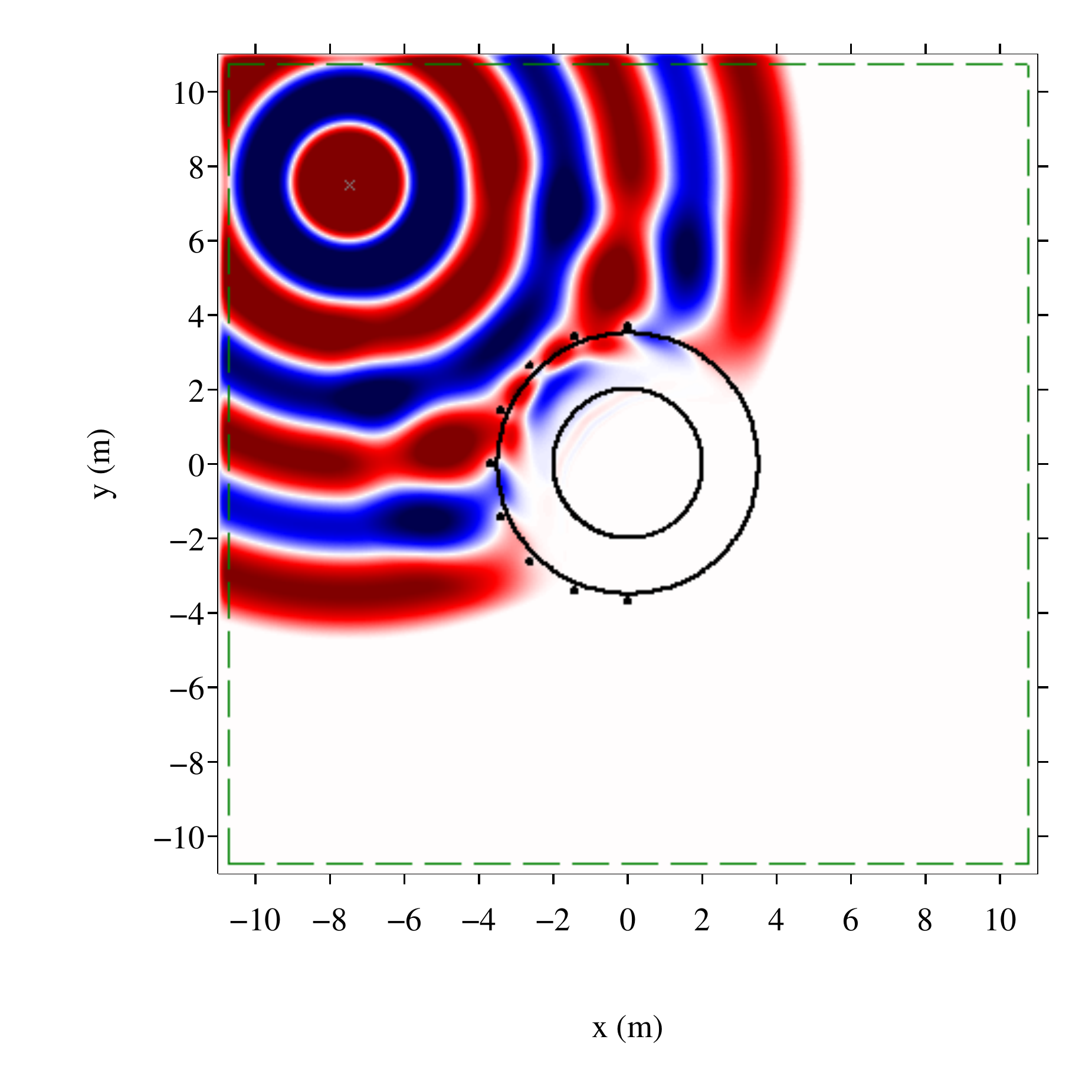}}
\subfloat[$t=71.2$\,ms]{\includegraphics[scale=0.25]{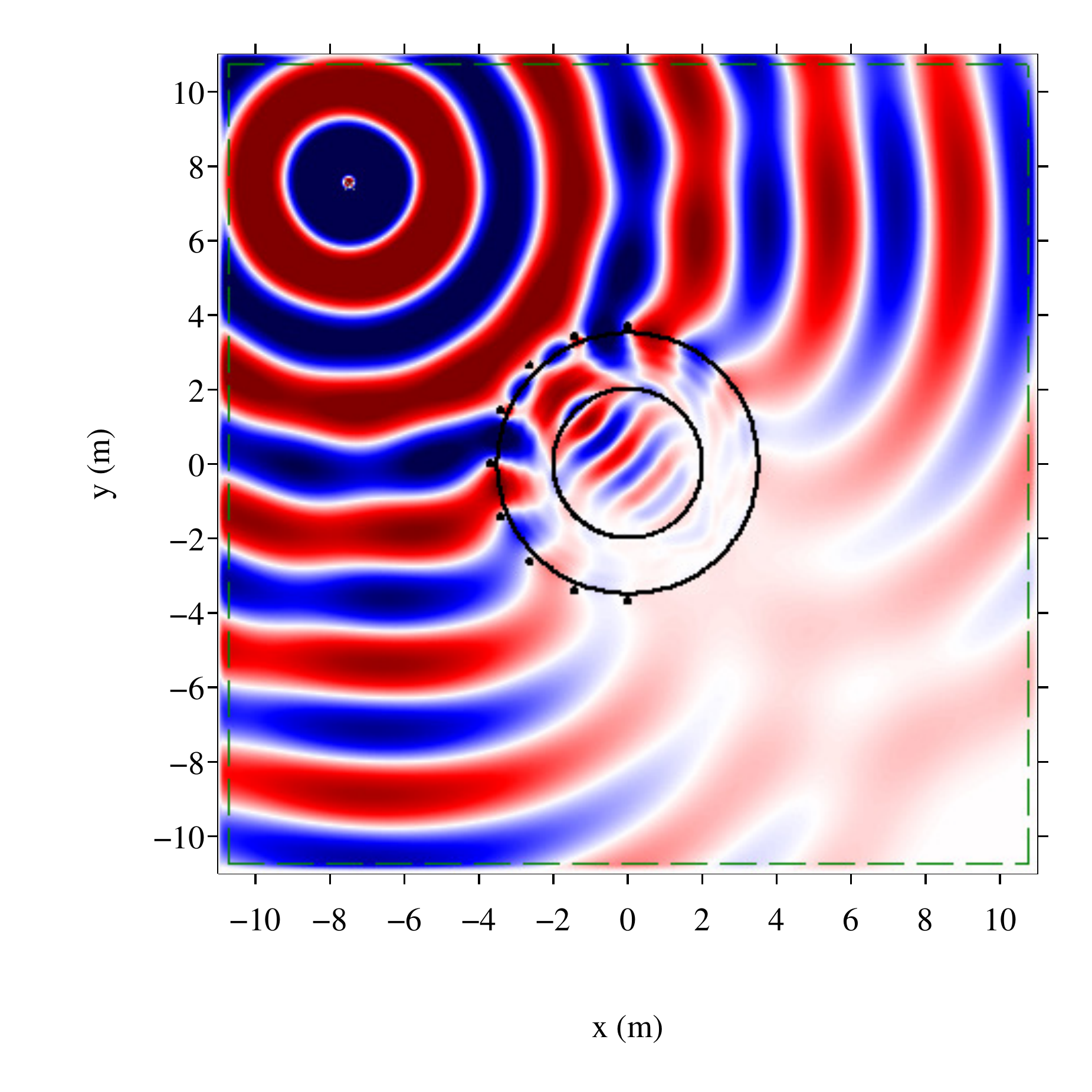}}\\[-4mm]
\subfloat[$t=1.424$\,s]{\hspace{-5mm}\includegraphics[scale=0.25]{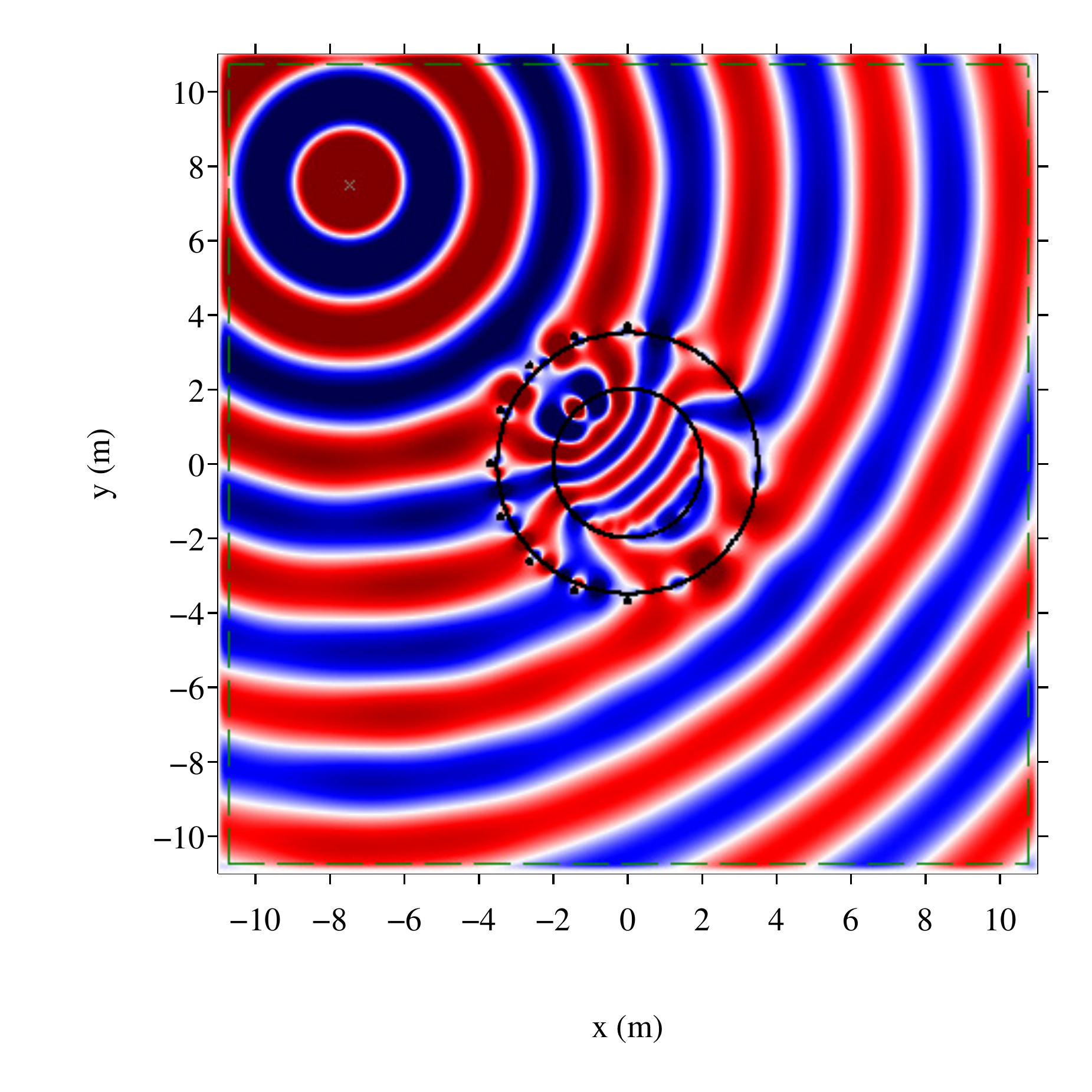}}
\subfloat[$t=1.424$\,s close-up]{\includegraphics[scale=0.25]{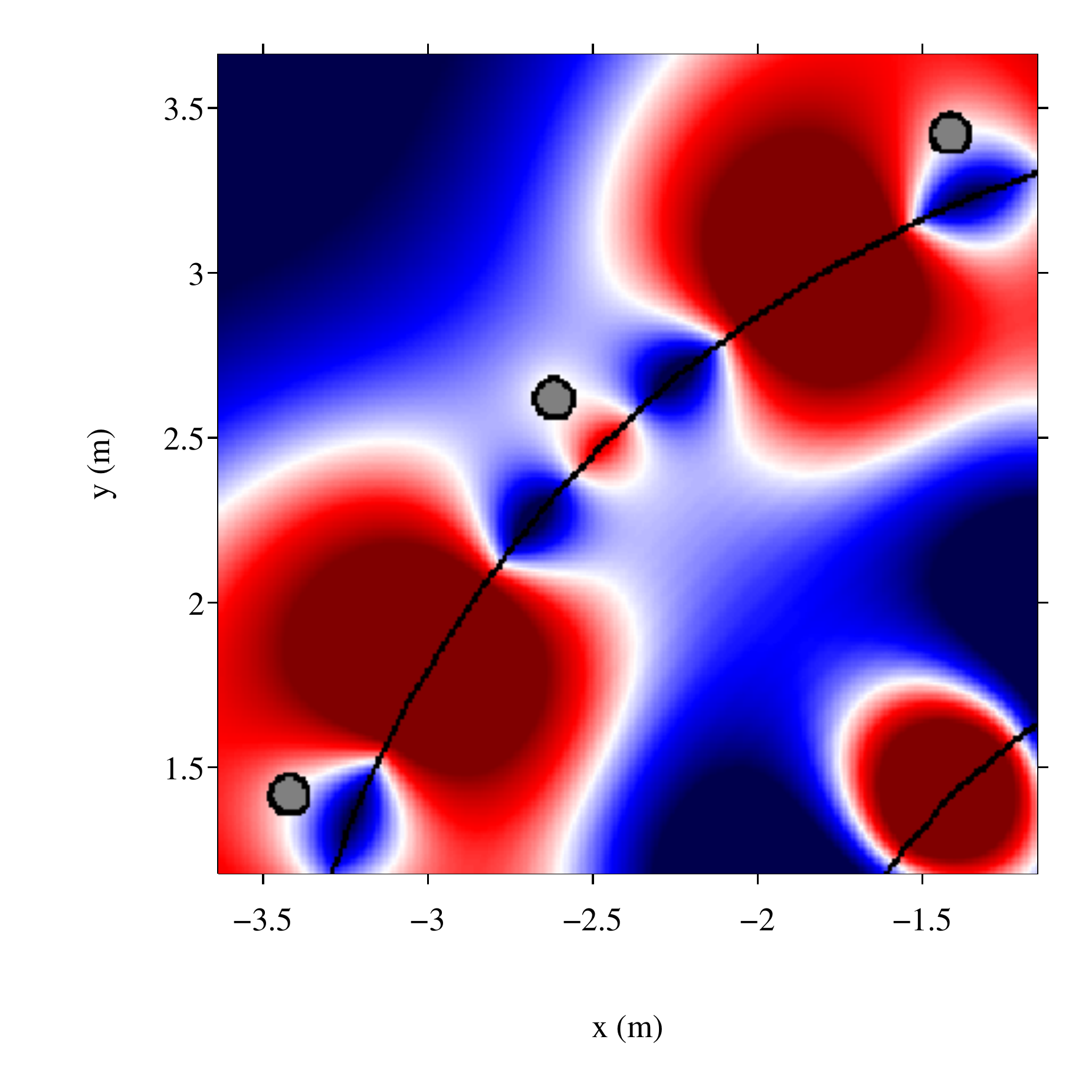}}
\vspace{-2mm}
\caption{\label{FigVar}Snapshots of the pressure field $p$ at different times {for scatterers with the cloak}. Panel (d) is a close-up of (c) highlighting the interactions of the surface waves on the shell with the scatterers placed in $\mathcal{F}$ (grey dots).}
\end{figure} 

Some scatterers are positioned in the cloaking region $\mathcal{F}$. These are nine equidistant and identical circular holes, with homogeneous Dirichlet condition $p=0$, whose centre is at a distance $d$ from the centre of the cloak and whose radius is $R$. Parameters are chosen as:
\[
\begin{aligned}
&R_c=2\mbox{\,m}, &&R_s=3.5 \mbox{\,m}, \\
&d=3.7 \mbox{\,m}, &&R=0.0583 \mbox{\,m},\\
&\rho_e=1.34 \mbox{\,kg.m}^{-3}, &&\kappa_e=1.34\cdot 345^2\mbox{\,Pa}.
\end{aligned}
\]
The cloaking region $\mathcal{F}$ is thus an annulus of width $R_*=6.125$\,m. The scatterers are put with a uniform polar angle $\theta$, from $\theta=\pi/2$ to $3\pi/2$, and they are illuminated by a source at point $\bm{x}_s=(-7.5,7.5)$\,m that emits a monochromatic signal at the central frequency $f_0=\omega_0/2\pi=100$\,Hz, switched on at $t=0$.
  
The systems \eqref{wave1} and \eqref{Auxiliary} are solved on a Cartesian computational grid with $N_1 \times N_2$ points. The grid is deemed fine enough with $N_1=N_2=1700$ to ensure a sufficient number of points discretizing the scatterers. A 4-th order space-time ADER scheme is used, together with a splitting algorithm. The discretization of the curved interfaces on the Cartesian grid considered requires care to avoid spurious diffractions and to accurately account for jump conditions. For this purpose, the Explicit Simplified Interface Method is used \cite{Lombard17}. Perfectly-matched layers are used on the external boundaries of the computational domain to simulate outgoing waves. Computations are performed using the software PROSPERO ({{\tt http://prospero-software.science/}}). All wave patterns shown in the sequel use the same linear color scale (ranging from dark blue for minimal values of field to dark red for its maximum values). 

\begin{figure}[h!]
\includegraphics[width=0.49\textwidth]{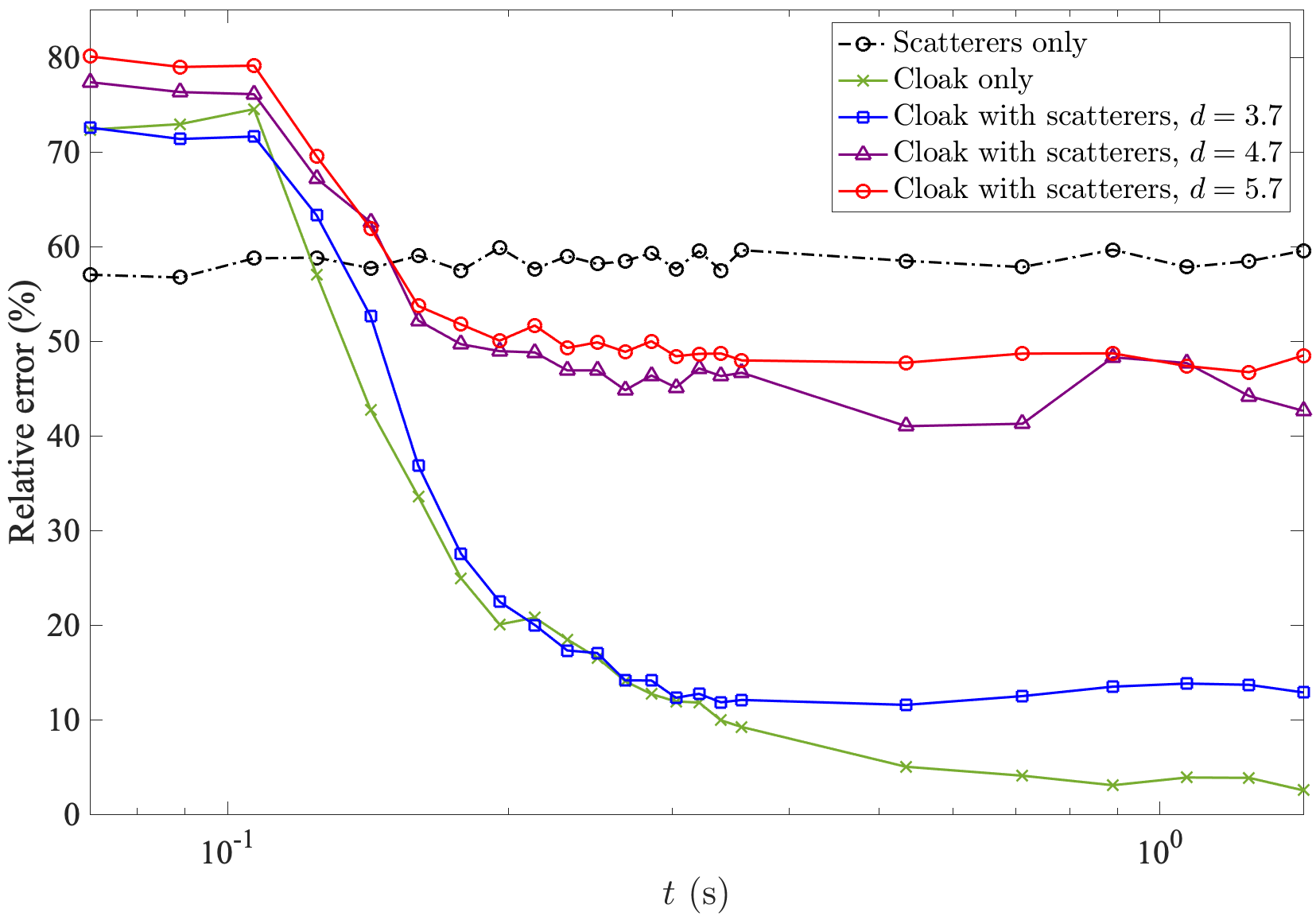}
\vspace{-0.5cm}
\caption{\label{FigErreurVsT}$L^2$-norm of the scattered pressure field in $\Omega_\text{obs}$ as a function of time (in log scale) and for various distances $d$ (in meters) to the cloak center.}
\end{figure}

In our first numerical experiment, we consider the cloak only, i.e. without scatterers. Figure \ref{FigCloak} illustrates the interaction of the transient waves with the cloak at different time steps. After a transient period (a-b), cloaking takes place (c). The scattered field measurements shown in Figure \ref{FigErreurVsT} confirm that cloaking is almost perfect. In (d), a close-up shows waves inside the shell. The effect of space folding can be observed.
{Dynamic animations
show {the propagation of negatively refracted waves} 
within the shell $\mathcal{S}$, where phase and group velocity have an opposite sign since the density $\rho$ and bulk modulus $\kappa$ are simultaneously negative (albeit being spatially varying) according to (\ref{ParamR}). However, within the core $\mathcal{C}$, $\rho$ and $\kappa$ are both positive constant, but the refractive index therein is larger than in the medium surrounding the cloak, which leads to an enhanced power flow in $\mathcal{C}$, similar to that in concentrators \cite{rahm08}. Finally, anomalous resonances at the core-shell ($|\bm{x}|=R_c$) and shell-outer medium ($|\bm{x}|=R_s$) interfaces of such a core-shell system predicted 25 years ago for an electrostatic configuration in \cite{oref1} manifest here a wave character since the condition for existence of surface waves akin to those already studied in the context of impedance-matched and sign-shifting electromagnetic media \cite{pendry02,guenneau05,bonnet10}, is naturally satisfied at all times by the principle of space-folding.}

\begin{figure}[htbp]
\centering
\subfloat[{Scatterers only}]{\hspace{-5mm}\includegraphics[scale=0.25]{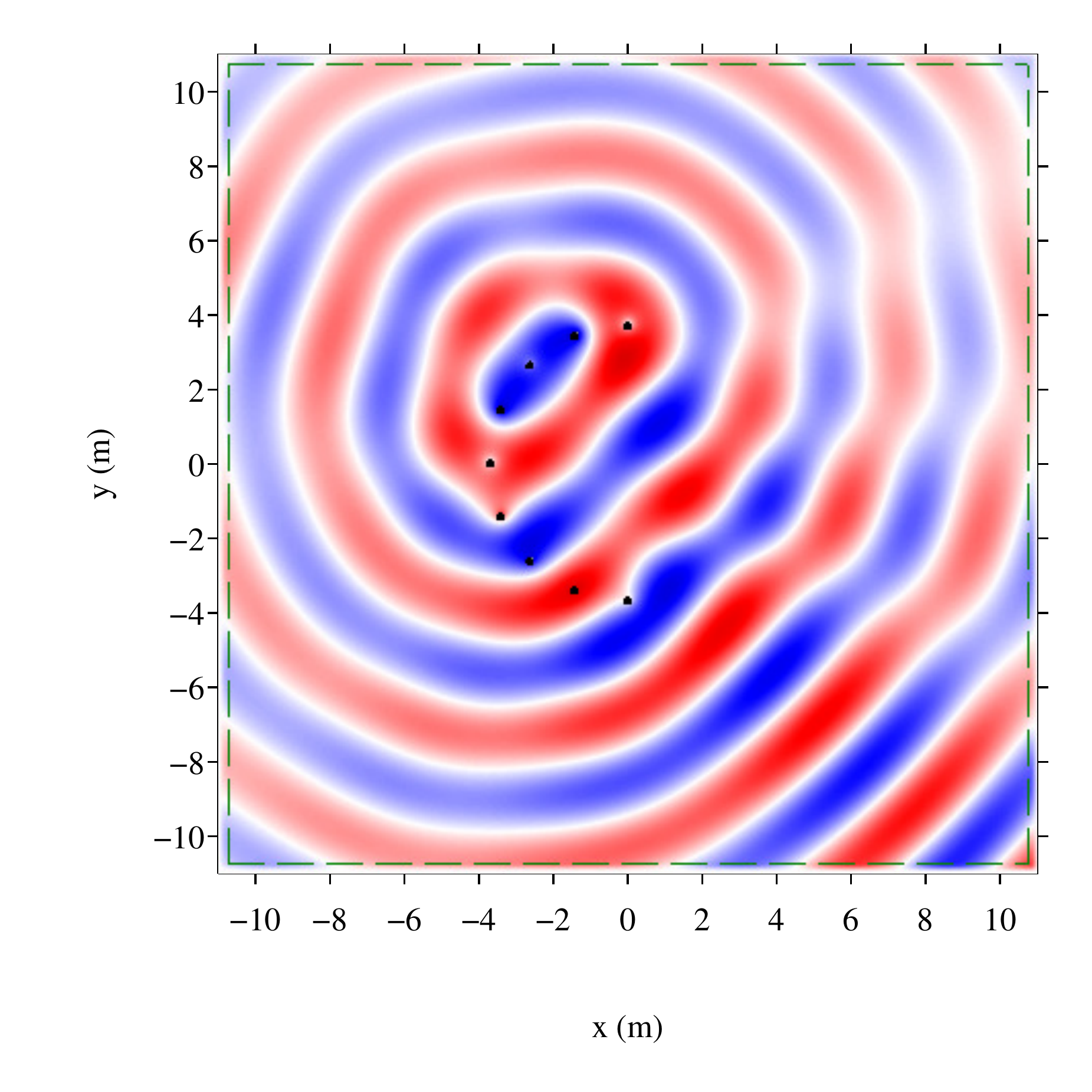}}
\subfloat[Scatterers and cloak]{\includegraphics[scale=0.25]{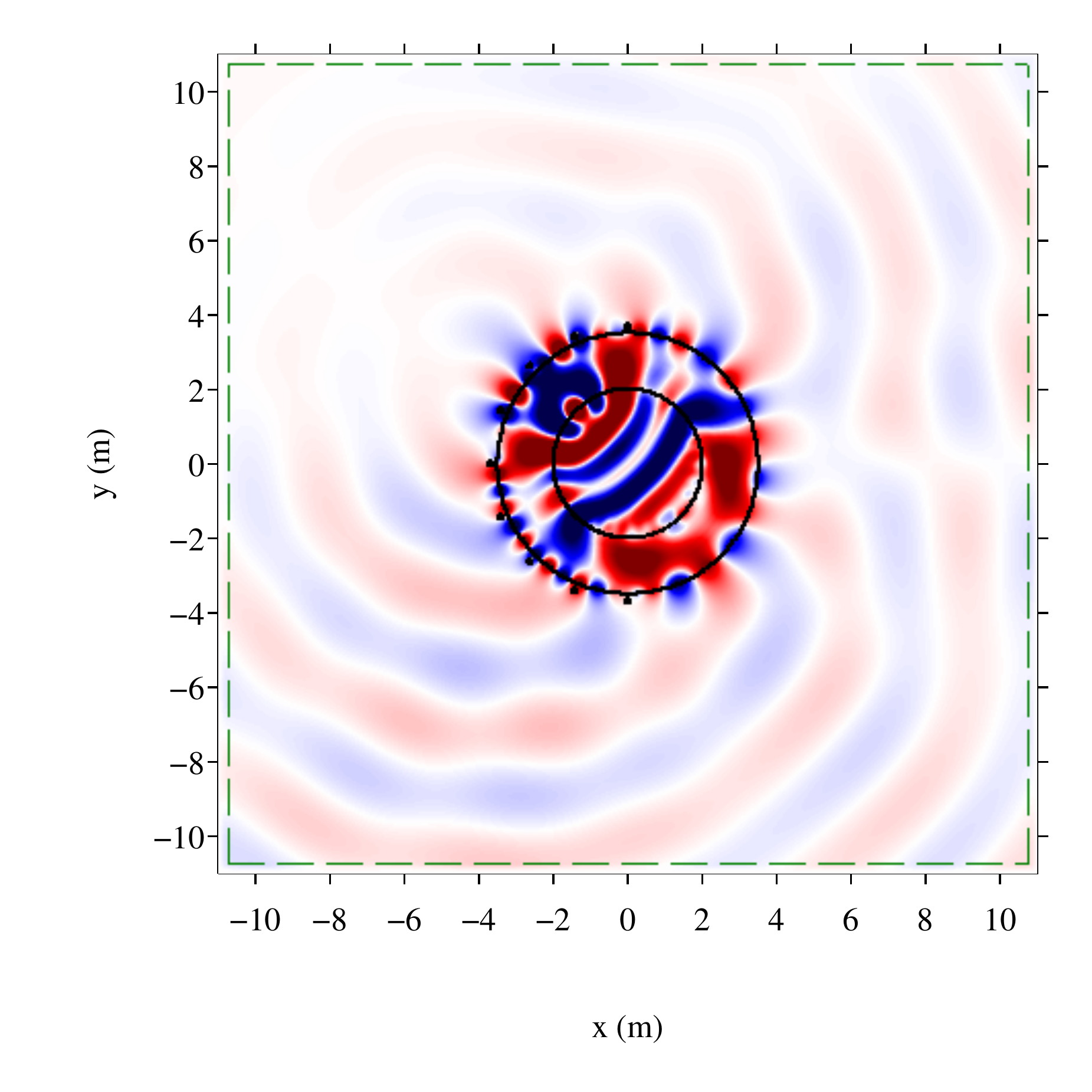}}\\[-4mm]
\subfloat[Same as (b), with $d=4.7$\,m]{\hspace{-5mm}\includegraphics[scale=0.25]{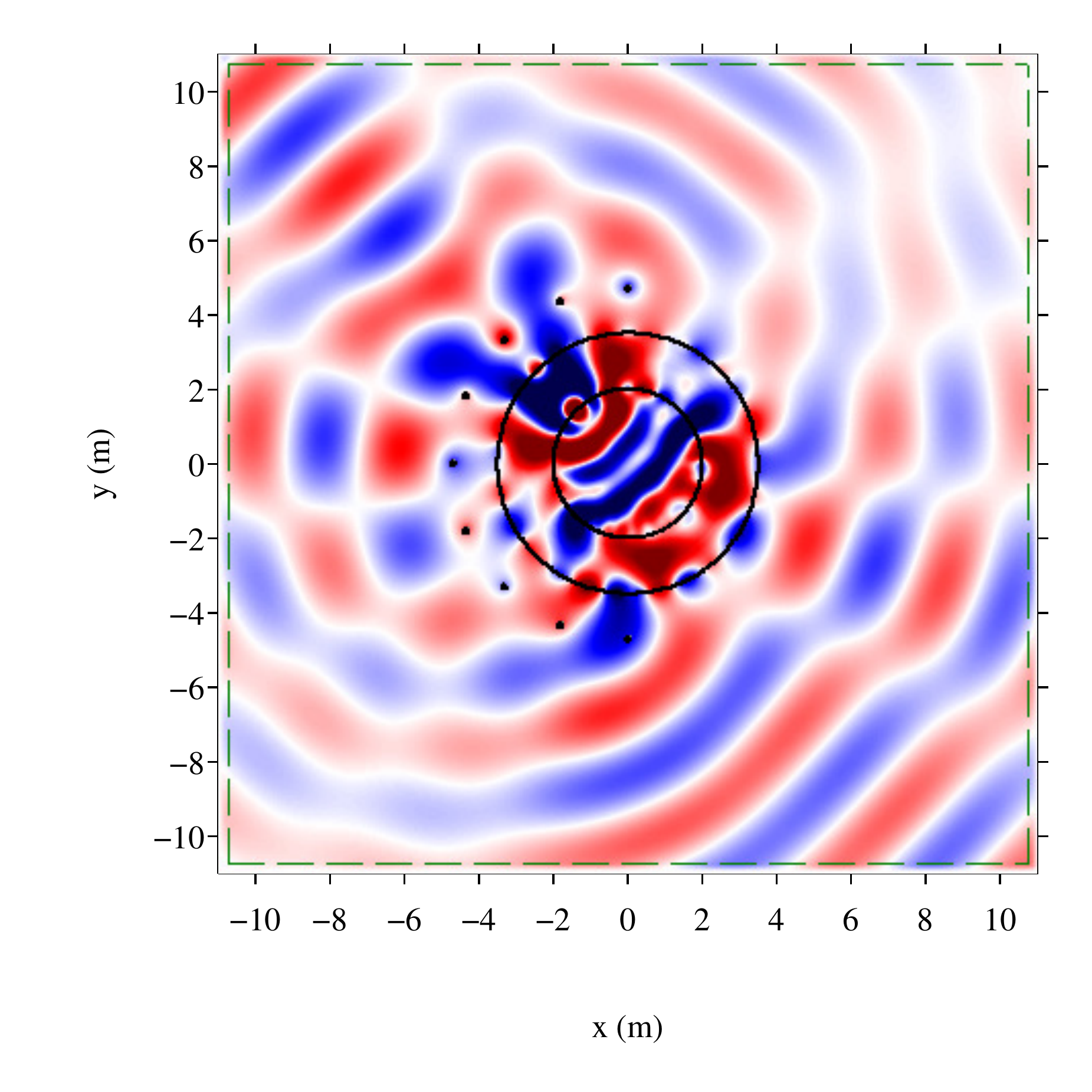}}
\subfloat[Same as (c), with $R=0.3$\,m]{\includegraphics[scale=0.25]{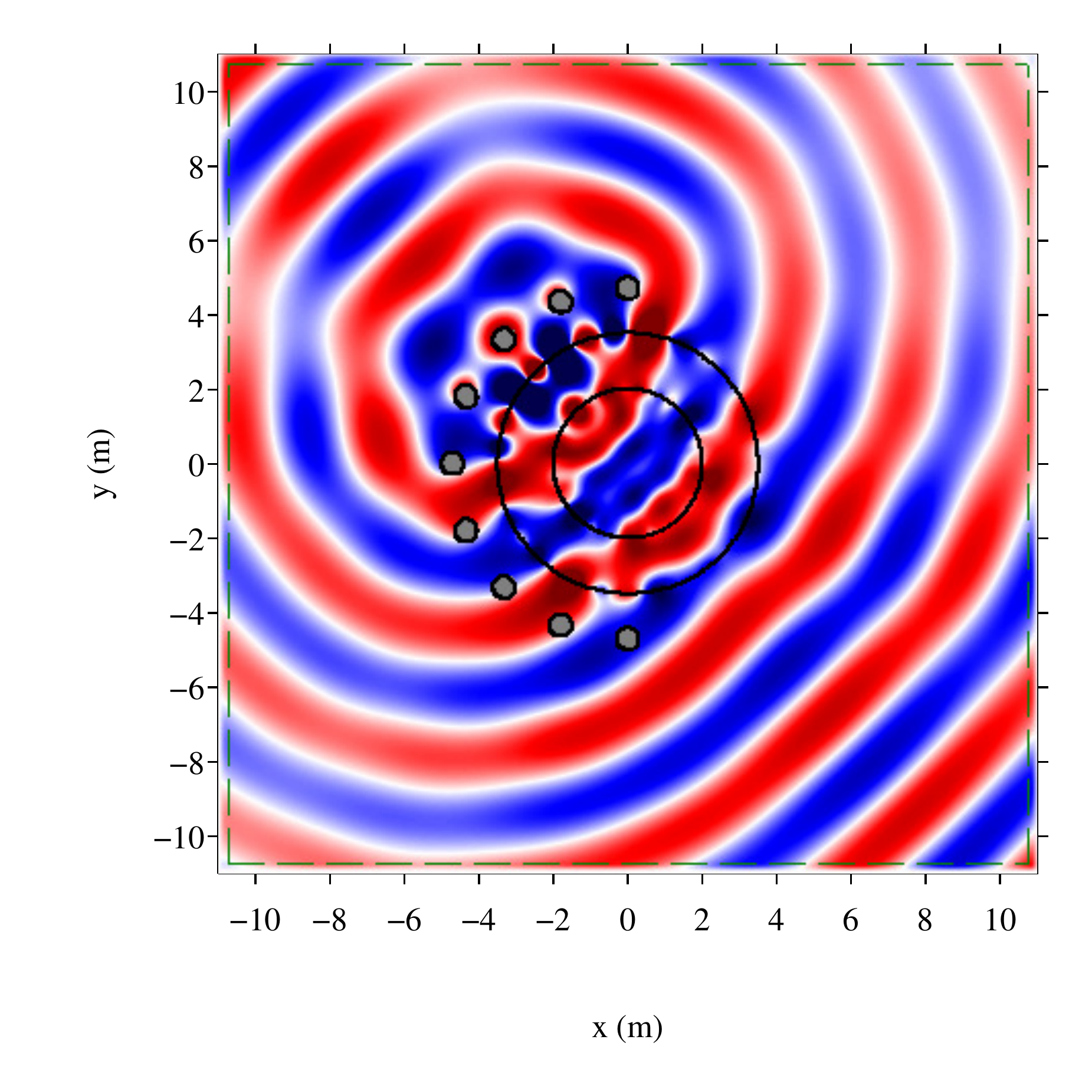}}
\vspace{-2mm}
\caption{\label{FigDiff}Snapshot of the scattered pressure field $p$ at time {$t=1.424$\,s}.}
\end{figure} 

Figure \ref{FigVar} illustrates the interaction of transient waves with the scatterers and the cloak, at different time steps. Clearly, at short times, the core-shell system does not cloak the scatterers, and behaves in a way similar to a superscatterer \cite{oref12}, whereas at longer times, when some permanent regime has been established, the diffraction by the scatterers is much reduced. A close-up on the scatterers in panel (d) shows the interaction of surface waves, at the matrix-shell interface, with the scatterers in the cloaking region $\mathcal{F}$
.

Figure \ref{FigDiff} displays snapshots of the scattered pressure field at $t=0.356$\,s in various configurations. The diffraction patterns of the scatterers (a) are significantly reduced by the cloak (b). In (c), the distance between the cloak and the scatterers is increased to $d=4.7$\,m: the interaction between surface waves and the scatterers is weaker, which degrades the efficiency of the cloak. This degradation is amplified when larger scatterers are considered, as in (d) with $R=0.3$\,m. These act somehow as a screen that prevents waves from reaching the cloak, so that surface waves cannot be generated to reduce scattering. One might argue that a possible route to achieve invisibility for larger scatterers, or even finite size possibly heterogeneous objects, that might be located further away from the cloak, would be to implement an external cloak based on complementary media as in \cite{lai09}. But this would imply that the design of the external cloak depends upon the objects to cloak, unlike in the present scheme. { We further note that one can in theory increase {\it ad libitum} the radius $R_*$ of the cloaking region by considering a vanishing radius $R_c$ for the core, which according to (\ref{ParamR}) requires a core of vanishing bulk modulus. This might open a route to external cloaking of finite size objects, or objects located further away from the cloak.}   

The $L^2$-norm $\|\cdot\|{=\sqrt{\int{\mid \cdot\mid}^2}}$ of the fields is measured in the rectangular domain $\Omega_\text{obs}=[-10,10]\times[-10,-7]$ opposite the source point. From now on, the relative error is defined as the ratio $\|p-p_\text{inc}\|/\|p_\text{inc}\|$, with $p$ being the solution to \eqref{wave1} and \eqref{Auxiliary} and $p_\text{inc}$ being the associated incident pressure field in the background medium. This relative error is reported in Figure~\ref{FigErreurVsT} and plotted as a function of time and for various distances $d$. Without the cloak, a steady state is reached rapidly and leads to a relative error around 60\,\%. With the cloak and scatterers at $d=3.7$\,m, the steady state is reached only at about 0.3\,s, i.e. 30 periods of the source. Once the steady state is reached, the external cloaking is all the more efficient than the scatterers are close to the core-shell system, as discussed previously: at $d=3.7$\,m, the relative error is around 12\,\%. For comparison, the cloak alone yields an error around 2.5\,\%.

\begin{figure}[h!]
\includegraphics[width=0.4\textwidth]{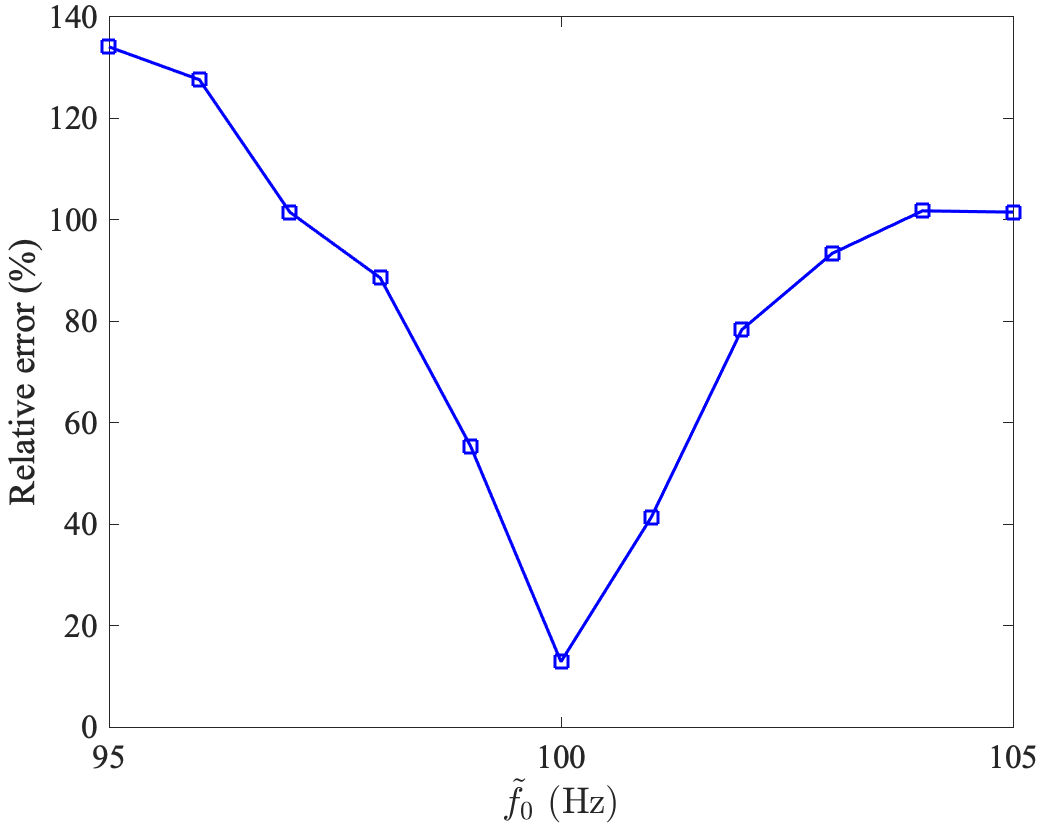}
\vspace{-0.5cm}
\caption{\label{FigErreurVsF}$L^2$-norm of the scattered pressure field {in $\Omega_\text{obs}$ as a function} of the source central frequency $\tilde{f}_0$.}
\end{figure}

To achieve the desired negative physical parameters in the shell, we emulate them as a realization of an effective Drude medium. The coefficient $\Omega_0$ in \eqref{Drude2} was defined from the frequency $f_0=100$\,Hz. Figure \ref{FigErreurVsF} then represents the relative errors  measured when the central frequency of the source is shifted as $\tilde{f}_0$. Cloaking performances appear to be significantly reduced when the incident frequency is not tuned so that properties \eqref{ParamR} are met within the shell, which highlights the sensitivity of cloaking to {the} physical parameters used: injecting $f=95$\,Hz in \eqref{Drude2} yields $g(0.95\,\omega_0)=-1.21$, which amounts to a 21\,\% error in the mass density in \eqref{ParamR}.

These efficiency measurements as a function of frequency raise two questions: (i) How to emulate negative parameters in the time-domain over a large frequency band? (ii) How to realize such a cloak in practice using a resonant micro-structured medium within the shell? These two questions, crucial to the design of external cloaking in the time-domain, will be the topic of future investigations.




\end{document}